\documentstyle[12pt]{article}
\begin{document}
\begin{center}
\baselineskip 2.0\baselineskip
{\large A new possibility to estimate the width, source location and velocity of halo CMEs. }

\vspace{10mm}

{\large G. Micha{\l}ek$^{1}$, N. Gopalswamy$^{2}$,  S. Yashiro$^{2}$,}
\end{center}

\parindent 0mm
{\small\em {$^{1}$Astronomical Observatory of Jagiellonian University, Cracow, Poland}}\\
{\small\em {$^{2}$Center for Solar and Space Weather, Catholic University of America,
\\ \hspace*{1cm}Washington, DC 20064}}\\

\begin{abstract}
It is well know that the coronagraphic observations of halo CMEs
are subject to projection effects.  Viewing in the plane of the
sky does not allow us to determine the crucial parameters defining
geoeffectivness of CMEs, such as the velocity, width or source
location. We assume that halo CMEs at the beginning phase of
propagation have  constant velocities, are symmetric and propagate
with constant angular widths. Using these approximations and
determining projected velocities and difference between times when
CME appears on the opposite sides of the occulting  disk we are
able to get the necessary parameters. We present consideration for
the whole halo CMEs from SOHO/LASCO catalog until the end of 2000.
We show that the halo CMEs are in average much more faster and
wider than the all CMEs from the SOHO/LASCO catalog.
\end{abstract}

\thispagestyle{empty}

\newpage
\baselineskip 1.0\baselineskip

\section{Introduction}
Space Weather is significantly controlled by coronal mass
ejections (CMEs) which can affect the Earth in a different way.
CMEs originating close to the central meridian, directed toward
the Earth, excite the biggest scientific concern. In coronagraphic
observations they appear as enhancement surrounding the entire
occulting disk and they were called `halo CME'. Since the first
identification by Howard et al. (1982) plenty of them were
detected and now they are routinely recorded by the high sensitive
SOHO/LASCO coronagraphs. In spite of large advantage over previous
instruments, the SOHO/LASCO observations are still affected by a
projection effect (Gopalswamy et al. 2000b). Viewing in the plane
of the sky does not allow us to determine the crucial parameters
defining geoeffectivness of CMEs, such as the velocity, width or
source location. Prediction of the arrival of CME in the vicinity
of Earth is critically important in space weather investigations.
Basing on interplanetary shocks detected by Wind and the
corresponding CMEs detected by SOHO, Gopalswamy et al. (2000a)
developed and next (Gopalswamy, 2001) improved an empirical model
to predict the arrival of CMEs at 1AU. The critical element
affecting this model is the initial CME speed. The better
prediction could be achieved if real initial velocities are used
instead projected velocities determined from LASCO observations.
Similarly, attempts made to estimate the projection effect based
on the location of the solar source employ ad hoc assumptions on
parameter such as the width of CMEs (Sheeley et al. 1999, Leblanc
and Dulk, 2000). In the present paper we try to determine these
crucial parameters defining geoeffectivness of CMEs, such as the
velocity, width or source location. We assume that halo CMEs at
the beginning phase of propagation have a  constant velocities,
are symmetric and propagate with  constant angular widths.
 Using these approximations and determining projected
velocities and difference between times when CME appears on the
opposite sides of the occulting disk we are able to get necessary
parameters. We present results for the whole halo CMEs from
SOHO/LASCO catalog until the and of 2000.

\section{The cone model of CME}
Typical limb CMEs observed by the LASCO coronagraph look like ejected blobs
of magnetized plasma with magnetic fields anchored to the sun surface.
This allow them to keep almost cone shape during  expansion through the C2 and
C3 fields of view. The observed angular widths, for many limb events, remain nearly
constant as a function of height (Webb et al. 1997, our founding
during work on SOHO/LASCO catalog). Most of them propagate with constant
radial frontal speed but many slow CMEs gradually accelerate and fast CMEs
decelerate (St. Cyr et al., 2000; Sheeley et al 1999., Yashiro et al., 2002).
Assuming that the halo CME propagate with a constant velocity and
angular width we can reproduce it by the cone model with four free parameters
such as the velocity, angular width and the orientation of the central axis.
 So we assume that bulk velocity of ejected blob is pointed radially and
isotropically. In the Fig.~1. schematically we show the basic
properties of our model. These assumptions should be true at the
beginning phase of CME expansion at least.  In the projection on
the symmetry plane it looks like a triangle represented by solid
thick arrows. The central axis of our CME is imaged by the dashed
thick arrow. Its inclination to the plane of sky is equal
$\gamma$. Each parts of this cone (triangle in projection) has a
constant real velocity $V$. The CME with the angular width
$\alpha$ is ejected from the Sun at distance $r$ from the central
meridian. Opposite parts of CMEs have velocities, projected on the
plane of sky,  equal respectively $\vec{V}x1$ and $\vec{V}x2$. In
the bottom of the picture we see the occulting disk of the
LASCO/C2 coronagraph. We have to note that it is only schematic
picture without a  real scale. We may observe that if CME
originates exactly in the center of the Sun it will appear at the
same time around the entire occulting disk. But if the source
location of CME is slightly shifted (=r)in  respect to the center
of the Sun (for example like in our picture) then CME will first
appear at the left (east) side of the occulting disk and finally
at the right (west) side of the occulting disk. Since this time we
can see in the LASCO picture full halo CME ring but slightly
asymmetric in respect to the occulting disk. The clue of this
method is based on this  asymmetry, in other words on  the
difference between times when CME appears at the opposite sides
(first and finally appearance) of the occulting disk. Considering
situation from our picture we can say that to see CME at the left
(east) side of the occulting disk it has to travel, in the plane
of sky, distance equal $2R-r$ with velocity $\vec{V}x1$. For this
work CME needs time $T_1$ equal
$$ T_1={(2R-r) \over \vec{V}x1}  \qquad .$$
Similarly, CME will appear on the right (west) side of the
occulting disk after time equal
$$ T_2={(2R+r) \over \vec{V}x2 }  \qquad.$$
From these equations we determine the difference time
$$ \Delta T=T_2-T_1={(2R+r) \over \vec{V}x2 }-{(2R-r) \over \vec{V}x1 }
\qquad .  \eqno(1)$$
From the geometry of CME shown in the Fig.~1. we get the rest necessary equations
$$cos(\gamma)={r \over R} \qquad , \eqno(2)$$
$$cos(\gamma-{\alpha \over 2})={Vx1 \over V} \qquad . \eqno(3)$$
$$cos(180^{o}-\gamma - {\alpha \over 2})= {Vx2 \over V} \qquad. \eqno(4)$$
We have four equations and four parameters to determine $r$, $\alpha$, $V$ and
$\gamma$. $Vx1$, $Vx2$ and $\Delta T$ we have to get from LASCO observations.

\subsection{Determination parameters describing halo CMEs}
Now we have to determine $Vx1$, $Vx2$ and $\Delta T$ from LASCO
observations. It is not so easy because typically the halo CMEs
are very faint and in addition their structure is very
complicated. We consider the whole halo events from SOHO/LASCO
catalog until the end of 2000. From LASCO observations we obtained
two height-time plots for each halo CME from our sample. The first
height-time plot is for this part of event which appears first
from the occulting disk. It is extrapolated to estimate time
($T1$) when this part of CME, in the plane of sky, reaches
heliocentric distance $=2R_{\odot}$ and to estimate velocity
$Vx1$. The second height-time plot is used to determine the same
parameters ($T2$ and $Vx2$) but  in the opposite side of the
occulting disk where the halo CME appears at last. An  example of
1999/06/28 CME observed by LASCO is present in the Fig.~2 (figure is to big to see it in this archive). In the
first panel at the time $T0$ we do not see any new event. In the
next panel, in the north-west
 quadrant of the Sun,  CME appears at time 07:31. From the height-time plot
we determine $T1=07:19$ and $Vx1=635km/s$. In the next panel, the
final part of CME appears in the south-west quadrant of the Sun.
From this part we determined $T2=07:34$ and $Vx2=515km/s$. In the
fourth panel we can see the full image of the halo CME at last.
The thick solid arrow present the axis along which the respective
parameters are determined. The position angle (PA = angle between
north pole of the sun and part of the halo CME where the Vx1 is
determined) is indicated also. So the difference time for this
event will be $\Delta T = T2-T1=15min$. Now from equations (1, 2,
3 and 4) describing our CME model we can determine  the $V$, width
and parameter $r$.

\section{Results}
 We made the same consideration for the rest
events from our sample. The results are present in the three
tables.
 Three first columns are got from the SOHO/LASCO catalog
(date, time and projected speed from LASCO observations). Next, in
four columns we have data received from our considerations of
LASCO images ($PA, Vx1, Vx2, \Delta T$). Parameters estimated from
our cone model $(r, \gamma, \alpha, V)$  are presented in columns
8,9,10 and 11. We also put, in  a column 12, a short
characteristic of a given event. Numbers from the range 0.0 until
3.0 describe quality of a given CME. The letter F informs that we
have frontside, B backside and B? probably backside halo CME. If a
halo CME is to faint to measure at list two hight-time points to
determine velocity we could not estimate necessary parameters so
we left empty space in our table and put quality $0$ in the column
12. Similarly, we could not determine the parameters for the
symmetric halo CMEs. This situation appears when asymmetry in
velocity is less than 10km/s minutes or in the difference time
less than 10 minutes. In this case we put `Sym' into column 12. In
column 13, if it is possible, we identified the source location
from GEOS X-flare onset.
\\
{\scriptsize
\begin{tabular}{|l||l||r||l||l||l||l||l||l||l||l||l||l|}
 \hline \hline
DATE& TIME & SPEED &PA& Vx1& Vx2 &$\Delta T$ &r& $\gamma$& $\alpha $& V &Char& Flare\\
 &  & ${km\over s}$& Deg &  ${km\over s}$ & ${km\over s}$  & Min &${1\over R_{\odot}}$ & Deg & Deg& ${km\over s}$& &    \\
\hline  \hline
1996/08/16&  14:14:06& 364&    96 &   405  &  220    &  62 &  0.17 &   80 &   59  & 660 & 1.0,B? & ---   \\
1996/11/07&  23:20:05&  497 &  114 &  412  &  361    &  18 &  0.16 &   80 &   133 & 429 & 1.0,B?& ---   \\
1996/12/02&  15:35:05&  538&   270 &  392  &  232    &  79 &  0.47 &   61 &   128 & 392 & 1.5,B? & ---   \\
1997/01/06&  15:10:42&  136&   182 &  100  &   85    &  75 &  0.13 &   82 &   105 & 117 & 0.5,F & S20W03   \\
1997/02/07&  00:30:05&  490&   260 &  297  &  160    &  140&  0.51 &   58 &   121 & 297 & 1.5,F  & S20W04  \\
1997/04/07&  14:27:44&  875&   126 &  956  &  551    &  23 &  0.42 &   65 &   139 & 954 & 2.0,F  &  S30E19\\
1997/05/12&  06:30:09&   464&  --- &  ---  &  ---    &  ---&  ---  &  --- &   --- & --- & Sym    &N21W08  \\
1997/07/30&  04:45:47&   104&  276 &   94  &   85    &  81 &  0.25 &   75 &   146 &  95 & 1.0,B  & ---  \\
1997/08/30&  01:30:35&   405&   65 &  397  &  163    & 103 &  0.21 &   78 &    56 & 590 & 1.0,F  & N30E17 \\
1997/09/28&  01:08:33&   359&   66 &   210 &  118    & 169 &  0.53 &   57 &  131  & 212 & 3.0,B  & ---  \\
1997/10/21&  18:03:45&   523&   30 &   527 &  356    &  35 &  0.24 &   75 &   103 & 580 & 1.0,F  & N20E12\\
1997/10/23&  11:26:50&   503&  --- &  ---  &  ---    &  ---&  ---  &  --- &   --- & --- & 0.0,B & ---  \\
1997/11/04&  06:10:05&   755&  --- &  ---  &  ---    &  ---&  ---  &  --- &   --- & --- & Sym    & S14W33 \\
1997/11/06&  12:10:41&  1556&  261 &  1524 &  765    &  34 &  0.82 &   34 &   153 &2059 & 1.5,F  & S18W63 \\
1997/11/17&  08:27:05&   611&  --- &  ---  &  ---    &  ---&  ---  &  --- &   --- & --- & Sym    & ---  \\
1997/12/18&  23:47:31&   417&   68 &   321 &   270   &  40 &  0.36 &   68 &   158 & 325 & 2.5,B  & ---  \\
1998/01/02&  23:28:20&   438&  258 &   281 &   142   & 197 &  0.93 &   20 &   165 & 602 & 2.0,B?  & ---  \\
1998/01/17&  04:09:20&   350&  --- &  ---  &  ---    &  ---&  ---  &  --- &   --- & --- & 0.0,B    & --- \\
1998/01/21&  06:37:25&   361&  176 &   387 &  265    &  80 &  0.71 &   44 &   159 & 468 & 0.5,F  & S57E19  \\
1998/01/25&  15:26:34&   693&   36 &   471 &  216    &  98 &  0.50 &   60 &   114 &  471& 1.0,F  & N24E27 \\
1998/03/29&  03:48:00&   1794& --- &  ---  &  ---    &  ---&  ---  &  --- &   --- & --- & 0.0,B?    & ---  \\
1998/03/31&  06:12:02&   1992& 167 &  1733 &   502   &  41 &  0.26 &   74 &    53 & 2591& 3.0,B? & ---  \\
1998/04/23&  06:55:20&   1618& 113 &  1744 &   945   &  21 &  0.51 &   59 &   126 & 1744& 3.0,F  & ---  \\
1998/04/27&  08:56:06&   1434&  --- &  ---  &  ---    &  ---&  ---  &  --- &   --- & --- & 0.0,F    & S16E50 \\
1998/04/29&  16:58:54&   1374&   16 &  1071 &  794    &  17 &  0.26 &    74&    111& 1134& 2.0,F  & S17E20  \\
1998/05/01&  23:40:09&    585&  142 &   623 &    367  &  31 &  0.1  &   84 &    40 & 1427& 2.0,F  &S18W05\\
1998/05/02&  05:31:56&    542&  143 &   661 &    426  &  23 &  0.1  &   85 &    39 & 1612& 2.0,F  &S20W17\\
1998/05/02&  14:06:12&    938&  --- &  ---  &  ---    &  ---&  ---  &  --- &   --- & --- & Sym    & S15W15 \\
1998/06/04&  02:04:45&   1802&  --- &  ---  &  ---    &  ---&  ---  &  --- &   --- & --- & 0.0,B    & ---  \\
1998/06/05&  12:01:53&    320&  223 &   170 &    109  &  215& 0.78  &   39 &   159 &  227& 1.0,F  & S23E43   \\
1998/06/07&  09:32:08&    794&  114 &  1117 &     834 &   17&  0.4  &   66 &   143 & 1122& 2.0,B  & ---  \\
1998/06/20&  18:20:37&    964&  153 &   964 &     481  &   54&  0.8  &   35 &   153 & 1285& 2.0,B? & ---  \\
1998/10/24&  02:18:05&    452&  116 &   404 &     377 &   32&  0.46 &   62 &   172 & 441 & 1.5,B? & ---  \\
1998/11/04&  04:54:07&    527&  0.0 &   390 &     158 &  114&  0.25 &   75 &    62 & 541 & 1.5,F  &N17W01\\
1998/11/05&  02:24:56&    577&  288 &   395 &     267 &   42&  0.18 &   79 &    88 & 482 & 1.0,F  & N19W10 \\
1998/11/05&  20:58:59&   1124&  305 &  1092 &     378  &  55 &  0.35 &   69 &    75 &1283 & 3.0,F  & N22W18\\
1998/11/24&  02:30:05&   1744&  224 &  1856  &    628&   43&  0.88 &   27 &   153 &2655 & 3.0,F & S30W81\\
1998/11/26&  03:42:05&    488&  --- & ---   &  ---    &  ---&  ---  &  --- &   --- & --- & 0.0    & ---  \\
1998/12/18&  18:21:50&   1745&   40 &   1758&     532 &   50&  0.68 &   47 &   120 & 1792& 2.0,F  & N19E64\\
1999/04/04&  04:30:07&   1178&  --- &  ---  &  ---    &  ---&  ---  &  --- &   --- & --- & 0.0,F    & N18E72 \\
1999/04/24&  13:31:15&   1495&  307 &   1259&     502 &  45 &  0.52 &   58 &   110 &1261 & 2.0,B  &--- \\
1999/05/03&  06:06:05&   1584&   50 &   1392&     345 &  61 &  0.61 &   51 &   110 &1369 & 2.0,F  &N15E32  \\

\hline \hline
\end{tabular}
\newpage
\begin{tabular}{|l||l||r||l||l||l||l||l||l||l||l||l||l|}
\hline \hline
DATE& TIME & SPEED &PA& Vx1& Vx2 &$\Delta T$ &r& $\gamma$& $\alpha $& V &Char& Flare\\
 &  & ${km\over s}$& Deg &  ${km\over s}$ & ${km\over s}$  & Min &${1\over R_{\odot}}$ & Deg & Deg& ${km\over s}$& &    \\
\hline  \hline
1999/05/10&  05:50:05&    920&   80 &  1080&       513&   33&  0.27 &   74 &    76 &1333 & 1.5,F  &N16E19  \\
1999/05/27&  11:06:05&   1691&  311 & 1700 &      623 &  42 &  0.71 &   44 &   130 &1821 & 1.5,B  &---  \\
1999/06/01&  19:37:35&   1772&  351 & 1792 &     662  &  32 &  0.40 &   65 &    88 &1902 & 1.5,B &---   \\
1999/06/04&  00:50:06&    803&    8 &   936&     475  &  38 &  0.37 &    68&   101 & 980 & 1.5,B? &---   \\
1999/06/08&  21:50:05&    726&   10 &    755&     690 &  19 &  0.49 &   60 &   170 & 834 & 1.5,F  &N30E03   \\
1999/06/12&  21:26:08&    465&  --- &  ---  &  ---    &  ---&  ---  &  --- &   --- & --- & Sym    & N22E37 \\
1999/06/22&  18:54:05&   1133&  --- &  ---  &  ---    &  ---&  ---  &  --- &   --- & --- & 0.0,F    & N22E37 \\
1999/06/23&  06:06:05&    450&  --- &  ---  &  ---    &  ---&  ---  &  --- &   --- & --- & Sym    & S10E71 \\
1999/06/23&  07:31:24&   1006&  --- &  ---  &  ---    &  ---&  ---  &  --- &   --- & --- & Sym    & S12E78 \\
1999/06/24&  13:31:24&    975&  --- &  ---  &  ---    &  ---&  ---  &  --- &   --- & --- & 0.0,F  & N29E13 \\
1999/06/26&  07:31:25&    558&    0 &   584 &     419 &  21 &  0.11 &   83 &    67 & 909 & 1.0,F  & N25E00\\
1999/06/28&  12:06:07&    560&  364 &  549  &    297  &  77 &  0.67 &   47 &   143 & 603 & 1.0,F & S27E55 \\
1999/06/28&  21:30:08&   1083&  --- &  ---  &  ---    &  ---&  ---  &  --- &   --- & --- & 0.0,F  & S25E49 \\
1999/06/29&  05:54:06&    589&  --- &  ---  &  ---    &  ---&  ---  &  --- &   --- & --- & 0.0    & ---  \\
1999/06/29&  07:31:26&    634&   10 &  635  &     515 &  15 &  0.15 &   81 &   112 &698  & 2.0,F & N18E07 \\
1999/06/29&  18:54:07&    438&  --- &  ---  &  ---    &  ---&  ---  &  --- &   --- & --- & 0.0,F  & S14E01  \\
1999/06/30&  04:30:05&   1049&  --- &  ---  &  ---    &  ---&  ---  &  --- &   --- & --- & 0.0    & ---  \\
1999/06/30&  11:54:07&    627&  193 &  588  &   424   &  23 &  0.16 &   80 &    92 & 705 & 1.0,F  & S15E00 \\
1999/06/30&  13:31:25&    514&  --- &  ---  &  ---    &  ---&  ---  &  --- &   --- & --- & 0.0    & ---  \\
1999/07/06&  17:06:05&    899&  350 & 1000  &   489   &  39 &  0.41 &   65 &   105 & 1026& 1.0,B & ---  \\
1999/07/19&  03:06:05&    509&  --- &  ---  &  ---    &  ---&  ---  &  --- &   --- & --- & 0.0,F    & N15W13  \\
1999/07/25&  13:31:21&   1389&  306 &1342   &  348    &  82 &  0.76 &   40 &   127 &1466 & 2.0,F  & N29W81 \\
1999/07/28&  05:30:05&    457&  --- & ---  &  ---    &  ---&  ---  &  --- &   --- & --- & 0.0,F   & S15E00 \\
1999/07/28&  09:06:05&    456&  --- & ---  &  ---    &  ---&  ---  &  --- &   --- & --- & 0.0,F   & S15E04 \\
1999/08/07&  23:50:05&    219&  --- & ---  &  ---    &  ---&  ---  &  --- &   --- & --- & 0.0,F   & S14E47  \\
1999/08/09&  03:26:05&    369&  --- & ---  &  ---    &  ---&  ---  &  --- &   --- & --- & 0.0,F  & S29W11 \\
1999/10/14&  09:26:05&   1250&   63 &  1362&    830  &  33 &  0.82 &   34 &   157 & 1899& 2.0,F  & N15E40 \\
1999/12/06&  09:30:08&    653&  154 &   680&    551  &  21 &  0.33 &   70 &   147 & 682 & 1.0,B?  & ---  \\
1999/12/12&  08:30:05&    720&  198 &  1118&   797   &   21&  0.50 &   59 &   147 & 1151& 1.0,B  & ---  \\
1999/12/20&  18:06:05&   1237&   15 &  1237&   783   &  23 &  0.28 &   73 &    74 & 2242& 2.0,B  & ---  \\
1999/12/22&  02:30:05&    482&   14 &   753&   525    &   42&  0.75 &  40  & 162   & 984 & 1.5,F  &N10E30\\
1999/12/22&  19:31:22&    605&   24 &   605&  515    &   44&  0.65 &  69  &  141  &1042 & 1.5,F  &N24E19\\
2000/01/14&  10:54:34&    229&  --- & ---  &  ---    &  ---&  ---  &  --- &   --- & --- & 0.0,B    & ---  \\
2000/01/18&  17:54:05&    739&  --- & ---  &  ---    &  ---&  ---  &  --- &   --- & --- & 0.0,F    & S19E11  \\
2000/01/25&  23:54:06&    222&  --- & ---  &  ---    &  ---&  ---  &  --- &   --- & --- & 0.0    & ---  \\
2000/01/27&  19:31:17&    828&  --- & ---  &  ---    &  ---&  ---  &  --- &   --- & --- & 0.0,F   & S09E71  \\
2000/01/28&  20:12:41&   1177&  --- & ---  &  ---    &  ---&  ---  &  --- &   --- & --- & 0.0,F    & S31W17 \\
2000/02/03&  12:30:05&    735&  --- & ---  &  ---    &  ---&  ---  &  --- &   --- & --- & 0.0,B  & ---  \\
2000/02/08&  09:30:05&   1079&   55 &  938 &  732    &    28& 0.63 &  50 &     162& 1091& 2.0,F  & N25E26 \\
2000/02/09&  19:54:17&    910&  218 &  1124 &  693    &    25& 0.44 &  63 &    128 &1125 & 1.5,F  & S17W40 \\
2000/02/11&  21:08:06&    498&   --- & ---  &  ---    &  ---&  ---  &  --- &   --- & --- & 0.0    & ---  \\
2000/02/12&  04:31:20&   1107&   --- & ---  &  ---    &  ---&  ---  &  --- &   --- & --- & 0.0,F    & N26W23  \\
2000/02/17&  20:06:05&    600&   196 & 660 &  540    &    23&  0.39 &  67  &    152&  668& 2.0,F  &S27W10 \\
2000/02/28&  10:54:05&    404&   279 & 466 &  370    &    43&  0.3  &  72  &    132&  475& 2.0,B? & ---\\

\hline \hline
\end{tabular}
\newpage
\begin{tabular}{|l||l||r||l||l||l||l||l||l||l||l||l||l|}
\hline \hline
DATE& TIME & SPEED &PA& Vx1& Vx2 &$\Delta T$ &r& $\gamma$& $\alpha $& V &Char& Flare\\
 &  & ${km\over s}$& Deg &  ${km\over s}$ & ${km\over s}$  & Min &${1\over R_{\odot}}$ & Deg & Deg& ${km\over s}$& &    \\
\hline  \hline
2000/03/01&  03:30:05&    529&   217 & 628 & 488     &    38&  0.64 &  49  &    162&  737& 2.0,B? & ---\\
2000/03/03&  05:30:07&    793&   --- & ---  &  ---    &  ---&  ---  &  --- &   --- & --- & 0.0,F  & S14W62  \\
2000/03/29&  10:54:30&    949&   --- & ---  &  ---    &  ---&  ---  &  --- &   --- & --- & 0.0,B    & ---  \\
2000/04/04&  16:32:37&   1188&   304 & 1281 &  641    &    40&  0.79 &   37 &   151 & 1645& 2.0,F  & N16W66\\
2000/04/10&  00:30:05&    383&   --- & ---  &  ---    &  ---&  ---  &  --- &   --- & --- & 0.0,F    & S14W01 \\
2000/04/23&  12:54:05&   1187&   279 &1309  &  533    &   46& 0.65  & 49   &   127 & 1351& 3.0,B  & ---  \\
2000/05/03&  02:06:05&    693&   --- & ---  &  ---    &  ---&  ---  &  --- &   --- & --- & Sym,B   & ---  \\
2000/05/05&  15:50:05&   1594&   269 & 1624 &  570    &   50& 0.85  & 32   &   146 & 2154& 2.0,F & S16W84\\
2000/05/12&  23:26:05&   2604&    63 & 2056&  699    &    36& 0.62  & 51   &   116 & 2072& 2.0,B & ?--- \\
2000/05/28&  11:06:05&    572&   --- & ---  &  ---    &  ---&  ---  &  --- &   --- & --- & 0.0,B?    & ---  \\
2000/06/02&  10:30:25&    442&  --- & ---   &  ---    &  ---&  ---  &  --- &   --- & --- & 0.0,F    & N10E23  \\
2000/06/06&  15:54:05&   1108&     6 & 1024 &  870    &    12& 0.32 &  71  &    152 & 1028& 2.5,F & N21E15 \\
2000/06/07&  16:30:05&    842&   --- & ---  &  ---    &  ---&  ---  &  --- &   --- & --- & 0.0,F  & N20E02 \\
2000/06/10&  17:08:05&   1108&   306 & 1376 &  710    &    32& 0.64  & 50   &    138&1460 &2.5,F  & N22W37\\
2000/07/07&  10:26:05&    453&   198 &  311 &  239    &    59& 0.42  &  65  &    147& 315 &1.5,B?  & ---  \\
2000/07/11&  13:27:23&   1078&    51 & 1453 &  1093   &    18& 0.68  &  47  &    162& 1753&2.0,F  & N18E27\\
2000/07/14&  10:54:07&   1674&   --- & ---  &  ---    &  ---&  ---  &  --- &   --- & --- & 0.0,F    & N22E07  \\
2000/07/27&  19:54:06&    905&   --- & ---  &  ---    &  ---&  ---  &  --- &   --- & --- & 0.0,F    & N10E07 \\
2000/08/09&  16:30:05&    702&   --- & ---  &  ---    &  ---&  ---  &  --- &   --- & --- & 0.0,F    & N11W09 \\
2000/09/12&  11:54:05&   1550&   216 & 1250 &    966  &    18& 0.58  & 54   &    159& 1385& 2.0,F    &S12W18\\
2000/09/12&  17:30:05&   1053&    47 & 1329&   681   &    27& 0.39  & 66   &    106& 1366& 2.0,B? & ---    \\
2000/09/15&  15:26:05&    481&   --- & ---  &  ---    &  ---&  ---  &  --- &   --- & --- & 0.0,F    &N14E02 \\
2000/09/15&  21:50:07&    257&   --- & ---  &  ---    &  ---&  ---  &  --- &   --- & --- & 0.0,F    & N14E01 \\
2000/09/16&  05:18:14&   1251&    21 & 1256&    946  &    12& 0.27  & 74   &    126& 1278& 2.0,F  & N14E04\\
2000/09/25&  02:50:05&    587&   --- & ---  &  ---    &  ---&  ---  &  --- &   --- & --- & 0.0,F & N15W28 \\
2000/10/02&  03:50:05&    525&   144 &  577&    381  &    42& 0.41  & 65   &   131 & 578 & 1.0,F & S08E05 \\
2000/10/02&  20:26:05&    569&   --- & ---  &  ---    &  ---&  ---  &  --- &   --- & --- & 0.0,F    & S08E05  \\
2000/10/09&  23:50:05&    798&   --- & ---  &  ---    &  ---&  ---  &  --- &   --- & --- & 0.0,F    & N02W18 \\
2000/11/01&  16:26:08&    801&   --- & ---  &  ---    &  ---&  ---  &  --- &   --- & --- & 0.0,F    & S17E39  \\
2000/11/03&  18:26:06&    291&   --- & ---  &  ---    &  ---&  ---  &  --- &   --- & --- & 0.0,F  & N02W02 \\
2000/11/08&  04:50:23&    474&   236 &  622&    294  &    77& 0.6   & 53   &   128 & 634 & 1.0,F & N10W77 \\
2000/11/08&  23:06:05&   1345&   --- & ---  &  ---    &  ---&  ---  &  --- &   --- & --- & 0.0,F    &N05W75  \\
2000/11/15&  23:54:05&    826&   --- & ---  &  ---    &  ---&  ---  &  --- &   --- & --- & 0.0,B    & ---  \\
2000/11/23&  06:06:05&    492&   230 & 450 &   334   &    48&  0.49 &  60  &    150&  466& 1.0,F  & S22W33\\
2000/11/24&  05:30:05&   1074&   352 & 996 &   734   &    21&  0.45 &  62  &    147& 1013&1.5,F   & N22W02 \\
2000/11/24&  15:30:05&   1245&   324 & 1396&  841    &    17&  0.26 &  74  &     96& 1556&3.0,F   & N22W07 \\
2000/11/24&  22:06:05&   1005&   312 & 1105&  575    &    37&  0.56 &  55  &    130& 1122&2.0,F   & N21W14\\
2000/11/25&  01:31:58&   2519&    75 & 2434&  724    &   34 &  0.54 &  57  &    100& 2452&2.0,F   &  N07E50\\
2000/11/25&  09:30:17&    675&   --- & ---  &  ---    &  ---&  ---  &  --- &   --- & --- & 0.0,B?    & ---  \\
2000/11/25&  19:31:57&    671&   --- & ---  &  ---    &  ---&  ---  &  --- &   --- & --- & 0.0,F    & N20W23 \\
2000/11/26&  17:06:05&   1026&   283 & 1240&   785   &   25 & 0.58 &  54    &  144 & 1303& 2.0,F   & N18W38 \\
2000/12/06&  17:26:05&    413&    --- & ---  &  ---    &  ---&  ---  &  --- &   --- & --- & Sym,B    & ---  \\
2000/12/18&  11:50:05&    510&    --- & ---  &  ---    &  ---&  ---  &  --- &   --- & --- & 0.0,F    & N14E03  \\
2000/12/28&  12:06:05&    930&    --- & ---  &  ---    &  ---&  ---  &  --- &   --- & --- & Sym,B    & ---  \\
\hline \hline
\end{tabular}
}
\subsection{Properties of the halo CMEs}
In the three tables we present list of the halo CMEs covering
period of time from August~1996 until the end of 2000. We have to
note that not all halo CMEs look identical. We have to consider
two types of halo CMEs. First, the classical full halo CMEs which
appear to surround the occulting disk very fast in the C2 LASCO
coronagraph. Generally they originate from close the disk center.
 Second, the wide limb CMEs which surround the
entire occulting disk very late, often in the field of view of the
C3 LASCO coronagraph. Sometimes limb events appears as halo due to
deflections of pre-existing coronal structures by the fast CME. So
we have to be very careful to distinguish between a real halo CME
and a limb fast event deflecting coronal material. We were able to
determine the respective parameters for 73th CMEs from our sample.
The rest CMEs had to more complicated structures, were to faint or
symmetric and it was to difficult to accomplish necessary
measurements. In the three histograms (Fig.~3, Fig.~4, Fig.~5) we
present distribution of $V$, $\alpha$, and $\gamma$. Here it is
important to note the halo CMEs seem to be much wider and faster
than typical events taken from the SOHO/LASCO catalog (Yashiro et
al. 2002). The average wide of halo CME is approximately equal
120$^o$ (more than two times that the value received from the
SOHO/LASCO catalog). The narrowest CME has width equal $40^o$ and
the widest one has the cone angle as large as $172^o$. The average
speed of the halo CMEs is $1080km/s$ (abut two times more than the
one from SOHO/LASCO catalog). The slowest one achieves velocity
equal $95km/s$ when the fastest is ejected with velocity
$2590km/s$. Fig.~5 show that the halo CMEs originate close to the
sun center (with $\gamma \geq 60^o$ ) with maximum of distribution
around $\gamma=65^o$. We have to remember that in our
consideration the symmetric CMEs which start exactly from the sun
center are not included. If we consider them the maximum of
$\gamma $ distribution could be shifted to the central meridian.
In the Fig.~6 we present the sky plane speeds against corrected
(real) velocities. The solid line represents linear fit to the
data points. The inclination of the linear fit suggests that the
projection effect slightly increases with speed of CMEs. It is
clear that the projection effect is  important and in average the
corrected speeds are 20 percents larger than the velocities
measured in the plane of sky.

\section{Summary}
In this paper we present possibility to estimate the crucial
parameters determining geoeffectiveness of the halo CMEs.  The
clue of this method is based on the difference between times when
the halo CME appears at the opposite sides (first and finally
appearance) of the occulting disk. We considered the whole events
form SOHO/LASCO catalog until the end of 2000. We were able to
determine
 the real velocity, width and source location for 73th CMEs from our sample. Unfortunately, 58
events were symmetric or too faint to do necessary considerations. Results are listed in the three
successive tables. This list could be use for further statistical examination or to prediction
 of the arrival of CME in the vicinity of Earth. Presented results suggest that
 the halo CMEs represent a specific class of CMEs which are very wide and fast.
 Using our results we have to remember that the
 simple model has several shortcomings: (i) CMEs may be
 accelerating, moving with constant speed or decelerating at the
 beginning phase of propagation. This means the constant velocity
 we assumed may not hold. (ii) CMEs may expand in addition to
 radial motion. Then the measured sky-plane speed is a sum of the
 expansion speed and the projected radial speed. This also would
 imply that the CMEs may not be a rigid cone as we assumed (Gopalswamy et
 al. 2001)
 (iii) The cone symmetry also may not hold. CME originating from
 loop structure could be elongated. All these limits can be
 overcome by stereoscopic observations only. Unfortunately, at the present time they
 are not available yet. It is necessary to develop the model to get the better
 fit to observations. The first step to improve our model could be achieved by
 consideration of acceleration and expansion
 of CMEs.
 \\
\\
{\small \bf Acknowledgments}\\
{\scriptsize This paper was done  during work of Grzegorz Michalek at Center
 for Solar and Space Weather,
Catholic University of America in Washington.\\
In this paper we used data from SOHO/LASCO CME catalog. This CME catalog
is generated and maintained by the Center for Solar Physics and Space Weather,
 The Catholic University of America in cooperation with the Naval
Research Laboratory and NASA. SOHO is a project of international cooperation between ESA and NASA.\\
Work done by Grzegorz Michalek was partly supported by {\it Komitet Bada\'{n} Naukowych} through
the grant PB 258/P03/99/17.}
\vspace{10mm}
\section*{References}
\parskip=0pt
\parindent=1cm
Gopalswamy, N., et al., Geophys. Res. Lett., 27, 145, 2000a\\
Gopalswamy, N., et al., Geophys. Res. Lett., 27, 1427, 2000b\\
Gopalswamy, N., et al., J. Geophys. Res., 106, 292907, 2001\\
Gopalswamy, N., 2001 Coronal Mass ejection:Initiation and Detection?????\\
Howard R.A., et al., Astrophys. J., 263, L101, 1982\\
Leblanc, Y., Dulk, G.A., J. Geophys. Res., 106, 25301, 2001 \\
Sheeley, N.R., Jr., Walters, J.H., Wang, Y.-M., Howard, R.A., J. Geophys.\\
\hspace*{1cm} Res., 104, 24739, 1999\\
St. Cyr, O.C., et al., J. Geophys. Res., 105, 18169, 2000\\
Webb, D.F., et al., J. Geophys. Res., 102, 24161, 1997\\
Yashiro, S.,  et al.,in preparation, 2002\\

\begin{figure}
\vspace{15cm} \includegraphics{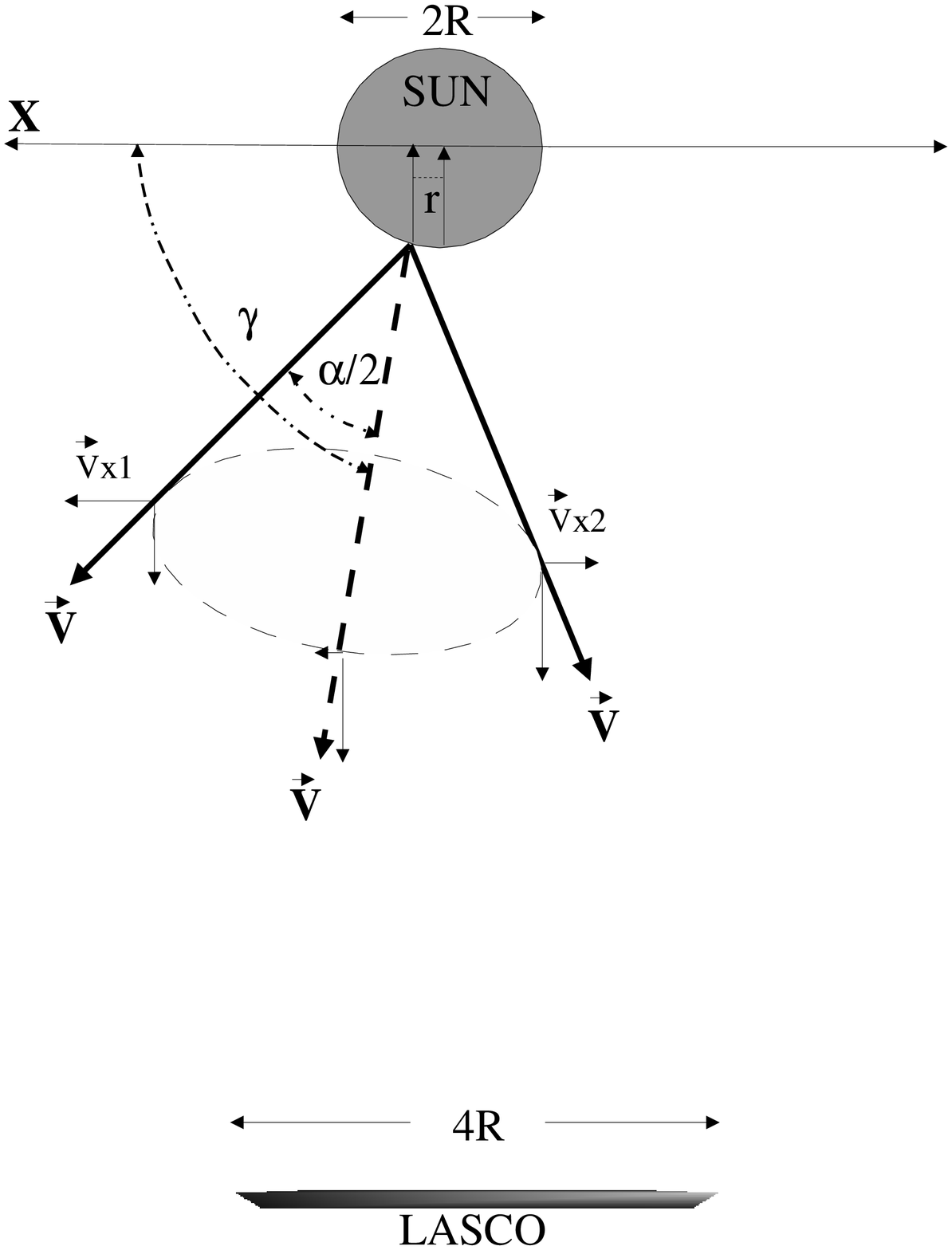} \vspace{0mm} \caption{The schematic
picture presenting our cone model of the halo CME.}
\end{figure}

\begin{figure}
\vspace{9cm} \includegraphics{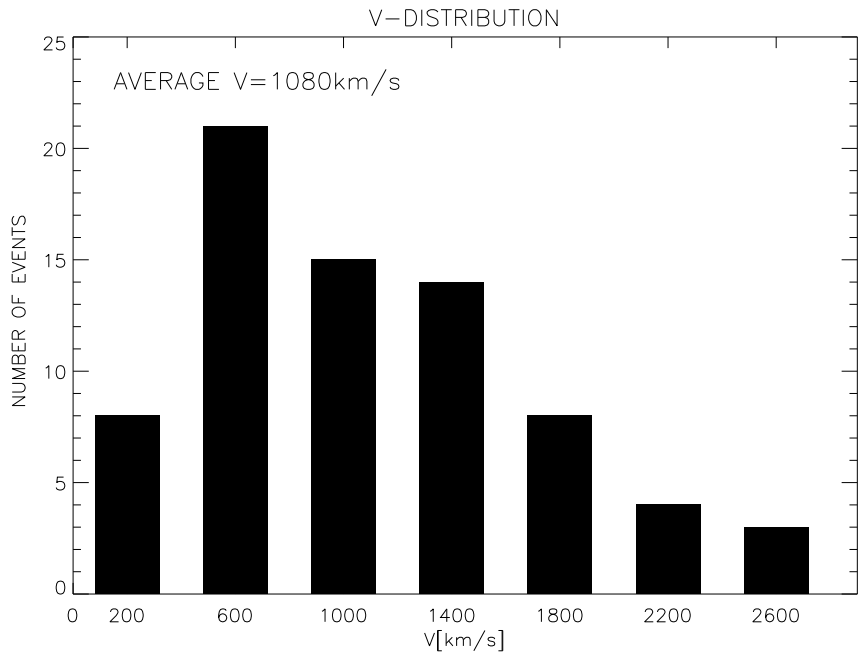} \vspace{0mm} \caption{The histogram
showing  distribution of $V$ for the halo CMEs.}
\end{figure}

\begin{figure}
\vspace{9cm} \includegraphics{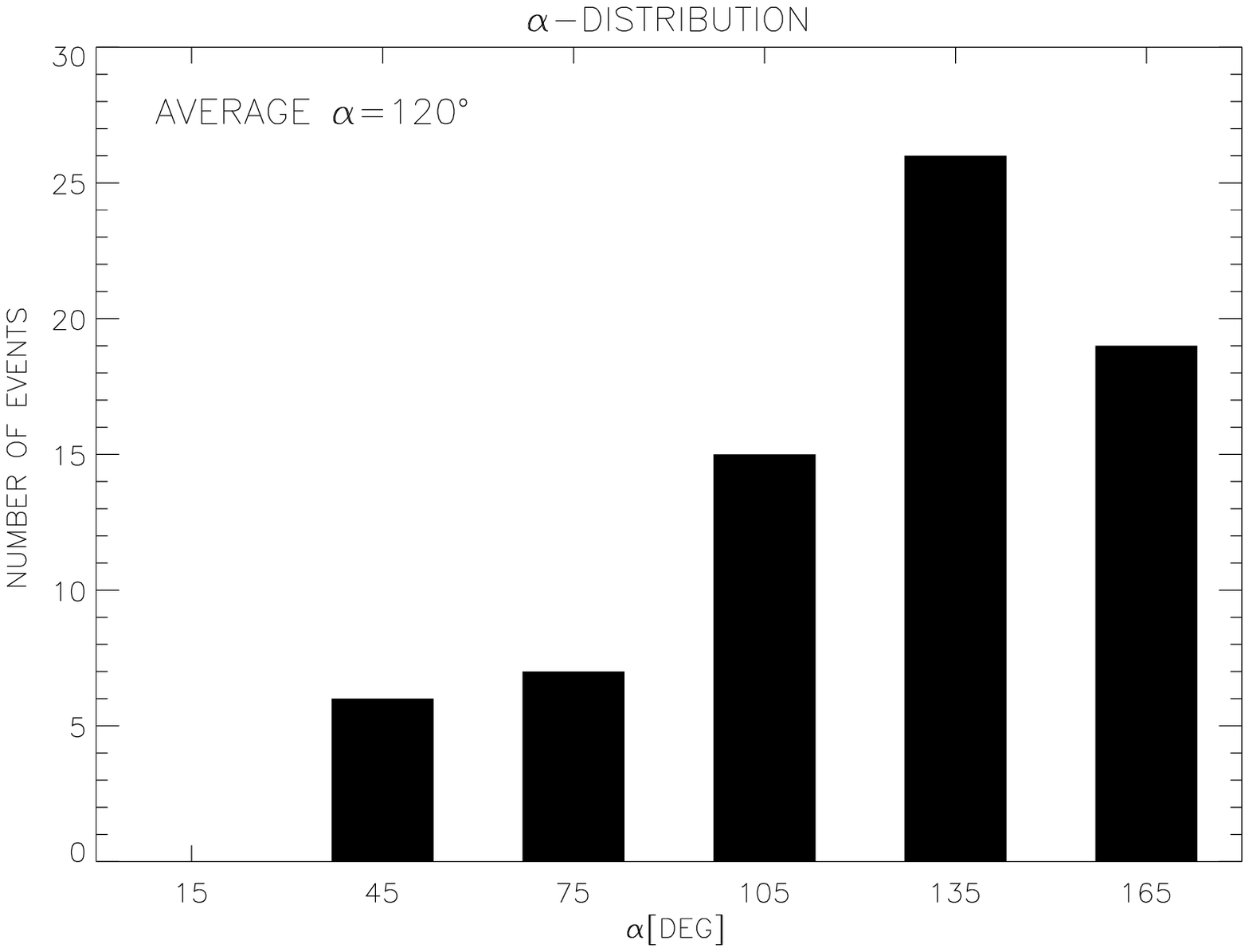} \vspace{0mm} \caption{The histogram
showing  distribution of $\alpha$ for the halo CMEs.}
\end{figure}

\begin{figure}
\vspace{9cm} \includegraphics{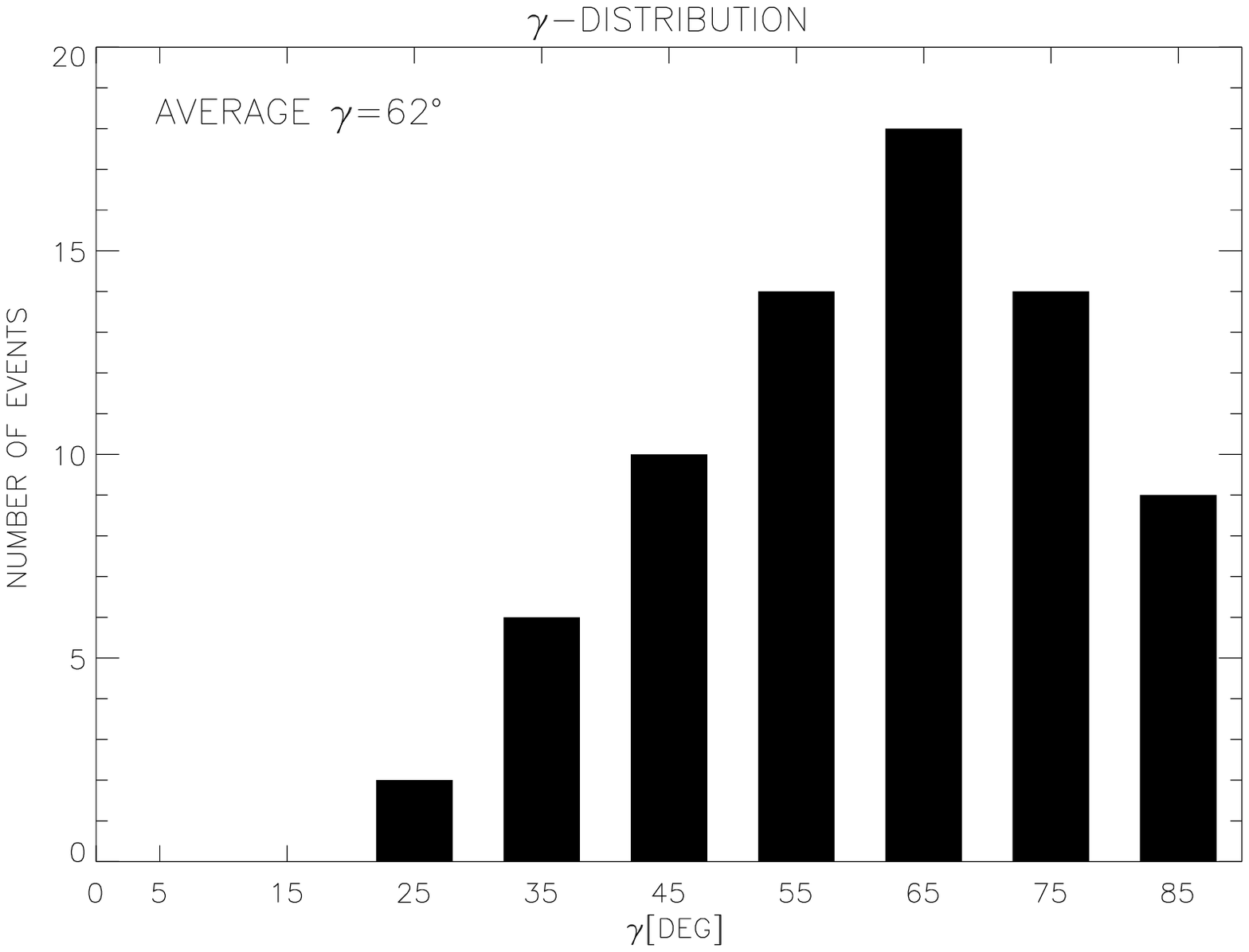} \vspace{0mm} \caption{The histogram
showing  distribution of $\gamma$ for the halo CMEs.}
\end{figure}

\begin{figure}
\vspace{9cm} \includegraphics{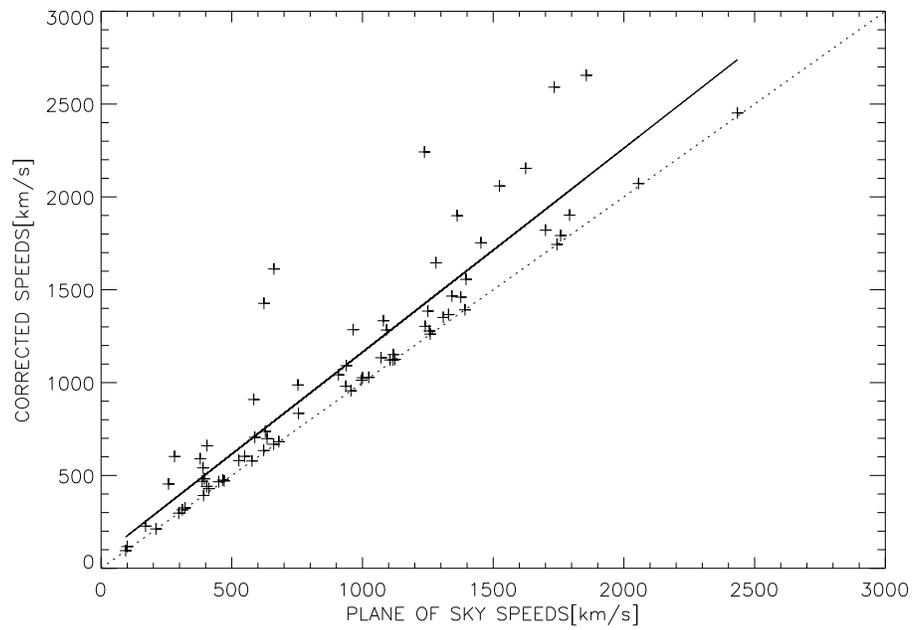} \vspace{0mm}\caption{The plane of the
sky speeds against the corrected (real) speeds. The solid line
shows the linear fit to data}
\end{figure}
\end{document}